\documentclass[aps,prc,twocolumn,floatfix,superscriptaddress]{revtex4-1}
\usepackage{graphicx,bm}
\usepackage{aas_macros}
\usepackage[colorlinks,citecolor=blue]{hyperref}
\begin{document}

\title{Neutrino nucleosynthesis: An overview}

\author{K. Langanke}
\affiliation{GSI Helmholtzzentrum f\"ur Schwerionenforschung,
  Planckstr. 1, Darmstadt, Germany}
\affiliation{Institut f{\"u}r Kernphysik (Theoriezentrum), Technische
  Universit\"at Darmstadt, Schlossgartenstr. 9, Darmstadt, Germany}
\author{G. Martinez-Pinedo}
\affiliation{GSI Helmholtzzentrum f\"ur Schwerionenforschung,
  Planckstr. 1, Darmstadt, Germany}
\affiliation{Institut f{\"u}r Kernphysik (Theoriezentrum), Technische
  Universit\"at Darmstadt, Schlossgartenstr. 9, Darmstadt, Germany}
\author{A. Sieverding}
\altaffiliation[Present address: ]{School of Physics and Astronomy, University of Minnesota, Minneapolis, MN 55455, USA}
\affiliation{Institut f{\"u}r Kernphysik (Theoriezentrum), Technische
  Universit\"at Darmstadt, Schlossgartenstr. 9, Darmstadt, Germany}
\affiliation{GSI Helmholtzzentrum f\"ur Schwerionenforschung,
  Planckstr. 1, Darmstadt, Germany}

\date{\today}

\begin{abstract}
  Neutrinos produced during a supernova explosion induce reactions on
  abundant nuclei in the outer stellar shells and contribute in this
  way to the synthesis of the elements in the Universe.  This neutrino
  nucleosynthesis process has been identified as an important
  contributor to the origin of $^7$Li, $^{11}$B, $^{19}$F, $^{138}$La,
  and $^{180}$Ta, but also to the long-lived radionuclides $^{22}$Na
  and $^{26}$Al, which are both key isotopes for $\gamma$-ray astronomy.  The
  manuscript summarizes the recent progress achieved in simulations of
  neutrino nucleosynthesis.
\end{abstract}	

\maketitle 

\section{Introduction}

Core-collapse supernovae play a crucial role for the origin of the
elements in the Universe 
\cite{Woosley.Heger.Weaver:2002}. 
First, and most importantly, the explosion frees 
nuclides, which have been produced by nuclear fusion reactions 
in the stellar interior during the star's long hydrostatic life, and mixes them
into the Interstellar Medium. Second, the explosion, related to
its hot and neutron-rich environment, can give rise to dedicated
nucleosynthesis processes like the $\gamma$-process, which produces
the heavy neutron-deficient isotopes \cite{Arnould.Goriely:2003}, 
or the rapid neutron capture
process ($r$-process) where modern supernova simulations imply
conditions supporting only the production of $r$-process nuclei up to
the A $\sim$130 mass region (weak~$r$-process \cite{Arcones.Montes:2011}, 
while the heavy $r$-process
nuclei are likely to be produced in neutron-star mergers 
\cite{Metzger.Martinez.ea:2010}. A unique feature
of core-collapse supernovae is the essential role played by neutrinos,
not only during the collapse phase (e.g. \cite{Bethe:1990,Langanke.Martinez.ea:2003}), 
but also for the explosion mechanism. 
The observation of neutrinos from SN1987a \cite{Hirata.Kajita.ea:1987,Bionta.Blewitt.ea:1987}
has confirmed that neutrinos are produced in overwhelmingly large numbers. Although the cross-sections for neutrino interactions 
with nuclei are tiny, they can therefore induce specific nucleosynthesis processes: 
The $\nu p$-process, that operates in the neutrino-driven wind just above 
a new-born neutron star \cite{Frohlich.Hauser.ea:2006,Pruet.Woosley.ea:2005,Frohlich.Martinez.ea:2006}, 
and the $\nu$-process, which is the focus
of this review.  

The $\nu$-process is initiated by neutrinos of all flavors produced in the hot and dense interior of the collapsed stellar
core that pass through the outer shells of the
star before and after the shock wave reaches this region. The neutrinos
interact with nuclei present in these shells by charged- (c.c.) and neutral-current (n.c.)
reactions. Here, it is important that
the energies of the neutrinos involved (up to about 20 MeV) restrict
c.c. reactions to electron neutrinos and electron antineutrinos,
while they are high enough to excite nuclei by n.c. reactions
to states above particle thresholds so that deexcitation proceeds via light particle
emission. Thus, both c.c. and n.c. reactions produce
new nuclides and, if these survive the fast reaction network
associated with the subsequent passage of the shock wave, contribute 
to nucleosynthesis. A handful of isotopes have been identified as being entirely or to
a significant portion created due to neutrino nucleosynthesis. Two important
radionuclides, $^{22}$Na and $^{26}$Al, are also affected.
As all neutrino flavors
contribute to the production of these nuclei, neutrino nucleosynthesis
serves as a test case for predicted supernova neutrino spectra and luminosities.  
 
Our review is organized as follows: in the next section we present
the input, strategies and outcomes of the neutrino nucleosynthesis
studies that were performed in the last three decades. \S~3 is devoted to 
a discussion regarding the sensitivity of the $\nu$~process to the neutrino spectra
and luminosities, while an outlook on future challenges and opportunities
is given in \S~4.

\section{Neutrino-nucleosynthesis studies}
\label{sec:nunucl}
The essential ingredients of neutrino nucleosynthesis studies are:
a) neutrino spectra and luminosities, b) neutrino-induced nuclear cross-sections, 
and c) the supernova model, including the stellar structure of the progenitor model as well as
effects of the shock wave. We briefly review these three items in turn.

a) Following the pioneering work by Woosley et
al. \cite{Woosley.Hartmann.ea:1990} and until recently, neutrino
spectra have been described by Fermi-Dirac distributions with
vanishing chemical potentials and time-independent temperatures
$T_\nu$ for the various neutrino flavors individually taken from
supernova simulations reflecting the cooling of the freshly born
neutron star mostly by neutrino pair emission. On general grounds,
simulations predict a hierarchy for the average neutrino energies:
$\langle E_{\nu_x} \rangle > \langle E_{\bar {\nu_e}} \rangle >
\langle E_{\nu_e} \rangle$.  (Here x refers to mu and tau neutrinos
and their antiparticles and the temperature is related to the average
energy by $T_\nu \approx 3.15\;\langle E_\nu \rangle $.)  With advanced
sophistication of supernova modeling, in particular by considering
additional processes affecting neutrino opacities, the values assumed
for the various neutrino temperatures (i.e. average energies) dropped
from $8\,\mathrm{MeV}$ to $4\,\mathrm{MeV}$ for $\mu$ and $\tau$
neutrinos, $6\,\mathrm{MeV}$ to $4\,\mathrm{MeV}$ for electron
antineutrinos and $5\,\mathrm{MeV}$ to $2.8\,\mathrm{MeV}$ for
electron neutrinos, where the higher values were adopted in the
pioneering work \cite{Woosley.Hartmann.ea:1990} and the lower values
in the recent study of Sieverding et al. \cite{Sieverding18}, taken
from the supernova simulation
\cite{Huedepohl.Mueller.ea:2010,Mirizzi.Tamborra.ea:2016}.  Heger et
al. \cite{Heger.Kolbe.ea:2005,Woosley.Heger:2007} used intermediate
values, which were appropriate at the time they performed their
studies: $T_{\nu_x} = 6\,\mathrm{MeV}$,
$T_{\bar{\nu}_e}= 5\,\mathrm{MeV}$ and $T_{\nu_e}= 4\,\mathrm{MeV}$.

Again following \cite{Woosley.Hartmann.ea:1990}, the neutrino luminosity has been assumed to
decrease exponentially with time, described by
\begin{equation}
L_\nu (t) = L_0 e^{ (- t/\tau)},
\label{eq:lum}
\end{equation}	
where $L_0 = 3\times 10^{53}\,\mathrm{erg}/\tau$ and $\tau=3\,\mathrm{s}$, consistent with simulations and observations.

b) Simulations of neutrino nucleosynthesis require
neutrino-induced partial nuclear cross-sections that take into account the decay
by particle emission (protons, neutrons, alpha). These cross-sections
are calculated applying a two-step procedure 
\cite{Kolbe.Langanke.ea:1992}. First, the neutrino-induced
cross-section leading to an excited nuclear state, via charged- or neutral current, 
is calculated
using a microscopic model, such as the Random Phase Approximation (RPA).
At the typical neutrino energies involved, Gamow-Teller and giant dipole
resonances dominate the cross-sections. Therefore, the RPA
is adequate because it describes the centroid and the total strength
of giant resonances quite well.
The decay of an excited state into different channels is calculated
employing the statistical model. RPA-based partial neutrino-nucleus 
cross-sections have been presented in 
\cite{Kolbe.Langanke.ea:1992,Cheoun.Ha.ea:2012,Balasi.Langanke.ea:2015,Huther:2014}, 
often already folded
with a neutrino energy distribution assumed appropriate for supernova
neutrinos. Ref. \cite{Sieverding18} has calculated 
partial neutrino-nucleus cross-sections 
spanning the entire nuclear chart and as a function of neutrino
energy, which allows us to study the influence of the dynamically changing
neutrino spectra on the neutrino nucleosynthesis process (see below).
For light nuclei, like $^4$He \cite{Gazit.Barnea:2007,Barnea.Gazit:2008,OConnor.Gazit.ea:2007} or $^{12}$C 
\cite{Suzuki.Chiba.ea:2006,Yoshida.Suzuki.ea:2008}, cross-sections calculated
with the hypersherical model or the interacting shell model exist, which both
are superior to RPA, because they also describe the fragmentation of the
giant resonance strength well. Cross-section relevant for the nucleosynthesis of
 $^{138}$La and $^{180}$Ta 
can be determined, at least to a large extent, from experimentally
determined Gamow-Teller strengths \cite{Byelikov.Adachi.ea:2007}.

c) As a secondary process, neutrino nucleosynthesis
predominantly operates on abundant seed nuclei and their presence
in the star determines where the process can operate. 
Thus, the final (one-dimensional) model of a star, 
calculated with a stellar evolution code, such as KEPLER \cite{Weaver.Zimmerman.ea:1978,Woosley.Heger:2007}, until 
just before collapse, is the starting point of a neutrino
nucleosynthesis study. Since self-consistent explosion models require large scale 
simulations \cite{Liebendoerfer.Mezzacappa.ea:2001,Janka.Hanke.ea:2012,Mueller:2016}, which are computationally too expensive for nucleosynthesis studies, the stellar model  is subjected to an artificial
explosion that usually employs a piston model tuned to a total energy of 
order $10^{51}$ erg \cite{Woosley.Weaver:1995}. The associated nucleosynthesis 
is studied with an extensive nuclear reaction network, 
including neutrino-induced reactions.
The neutrino spectra and luminosities are modeled as described above.

Starting with the pioneering work of Woosley et al. \cite{Woosley.Hartmann.ea:1990}, 
various studies
have identified a set of nuclei whose galactic abundance
is being produced dominantly or in significant portion by neutrino
nucleosynthesis: $^7$Li, $^{11}$B, $^{15}$N, $^{19}$F, $^{138}$La, $^{180}$Ta, 
and
the radionuclides $^{22}$Na and $^{26}$Al.

 \begin{table}
\caption{Production factors normalized to $^{16}\mathrm{O}$ of $\nu$-process
 isotopes for a $15\,\mathrm{M}_\odot$ progenitor model for a range of model assumptions about the
 neutrino emission spectra in recent studies. The reduction of the expected
 neutrino energies in modern supernova simulations has reduced the predicted
 production factors, whereas taking into account the full spectral information increases the yields. Note that the nuclear and neutrino reaction cross-sections have changed 
 from Heger et al. (2005) to Sieverding et al. (2019)}
 \label{tab:prodfac}
  \begin{ruledtabular}
  \begin{tabular}{lccc}
   Nucleus & High energies & Low energies & Simulation  \\
           & \tiny{Heger et al. (2005)} & \tiny{Sieverding et al. (2018)} &    \tiny{Sieverding et al. (2019)}   \\ \hline
 $^{7}$Li         &    --   &  0.083 &   0.187  \\
 $^{11}$B         &   1.884  &  0.280 &  0.516  \\
 $^{15}$N         &   0.487  &  0.116 &  0.141  \\
 $^{19}$F         &   0.602  &  0.180 &  0.209  \\
 $^{138}$La       &   0.974  &  0.487 &  0.824  \\
 $^{180}$Ta$^{m}$ &   0.964  &  0.484 &  0.636  \\
  \end{tabular}
\end{ruledtabular}
\end{table}

Table \ref{tab:prodfac}
gives the production factors for these nuclei relative to solar abundance \cite{Lodders:2003}
and normalized to $^{16}$O, obtained by
neutrino nucleosynthesis calculations for a star with an initial mass of $15\,\mathrm{M}_\odot$, representative for 
the nucleosynthesis of massive stars as explored by
\cite{Sieverding18} for a range of stellar models between $13\,\mathrm{M}_\odot$ and $30\,\mathrm{M}_\odot$.
This study, which for the first time considered a complete set of partial
nuclear reaction cross-sections, did not give evidence for other nuclei
being produced by neutrino nucleosynthesis. Two important
conclusions  can be derived from Table \ref{tab:prodfac}. First, neutrino nucleosynthesis
produces a large fraction of the solar yields for 
$^{11}$B, $^{138}$La and $^{180}$Ta,
and also contributes noticeably to the solar $^{15}$N and $^{19}$F 
abundances. But
for all of these nuclides other nucleosynthesis processes are required
to reproduce the full solar yields. Secondly, the effectiveness
of neutrino nucleosynthesis is severely reduced if modern supernova
neutrino spectra with lower average energies are considered rather
than those spectra appropriate some years ago, e.g. those used by Heger
et al. (2005) \cite{Heger.Kolbe.ea:2005}. 
The modern spectra reduce predominantly the neutral-current
contributions to the calculated yields as significantly less high-energy
neutrinos are available to excite the nucleus above particle thresholds. This
is particularly important for tightly bound nuclei like 
$^{12}$C, $^{16}$O and $^{20}$Ne which
have rather high thresholds for proton and neutron emission, that - as we
will see below - are important for the neutrino nucleosynthesis production mechanisms 
for $^{11}$B, $^{15}$N and $^{19}$F, respectively. 
As the c.c. cross-sections are,
in relative terms, less affected by the modifications of the supernova neutrino spectra,
the relative weight  
to the total yields is shifted towards c.c. contributions.

\begin{figure}
\includegraphics[width=\linewidth]{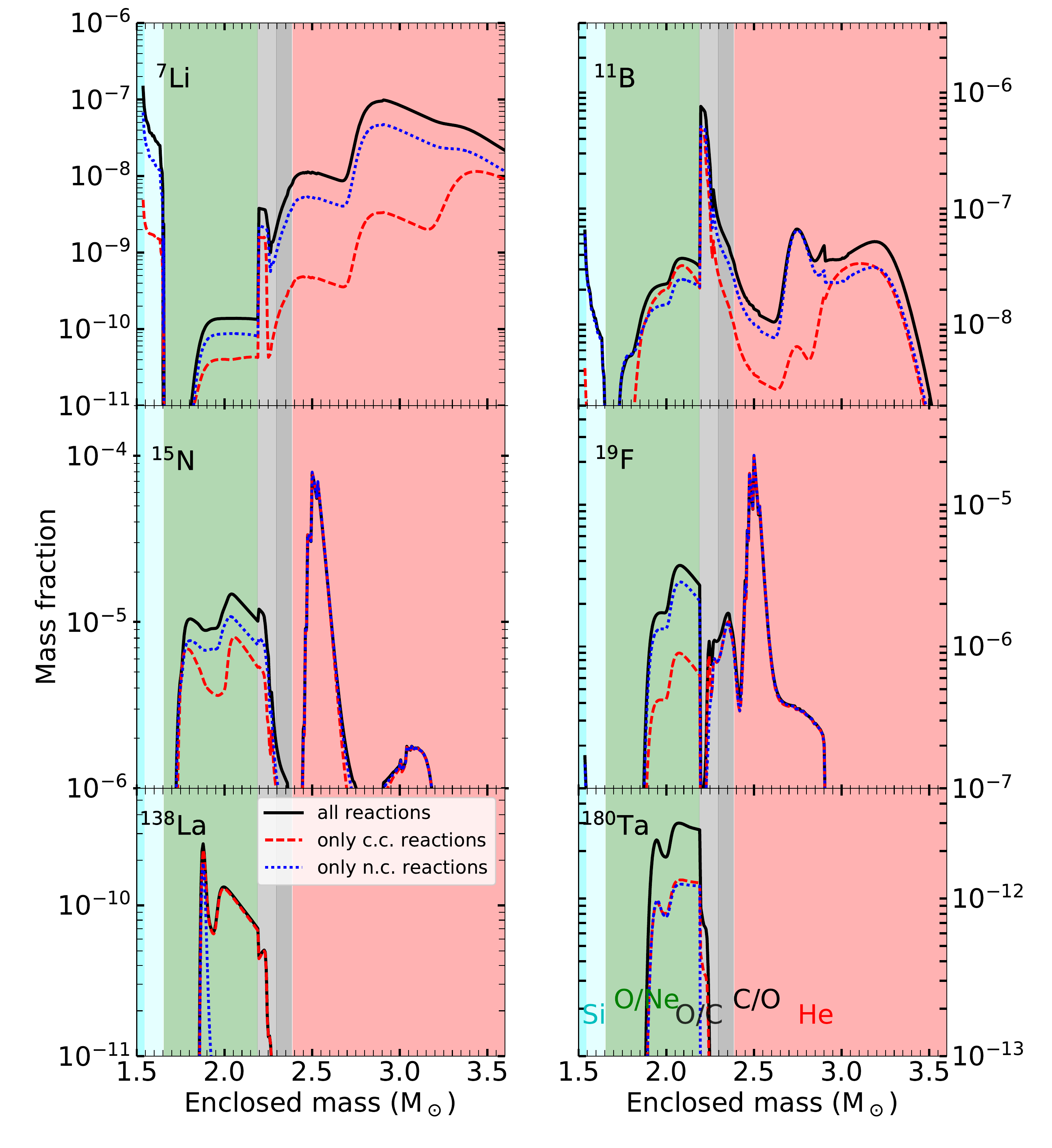}
\caption{Mass fraction profiles for the $15\,\mathrm{M}_\odot$
  progenitor model with low neutrino energies as chosen in
  Ref. \cite{Sieverding18} for the six isotopes most
  affected by the $\nu$-process. Red dashed and blue dotted lines
  indicate results with only charged current (c.c.) and only neutral
  current (n.c.) reactions. Note that the scale is different in each
  panel. The background colors indicate the different compositional
  layers of the stellar model, indicated in the bottom right
  panel.}
\label{fig:all_profile}
\end{figure}

In the following sections we will discuss  the production sites and reactions
for the different nuclides produced by the $\nu$-process.

For illustration, we use again the results from a calculation of a 15 solar mass star.
Fig. \ref{fig:all_profile} shows the mass fractions of the selected nuclides as function of
enclosed mass. The different compositional shells of the star that result from the hydrostatic burning stages
(He, C/O, O/C, O/Ne and Si) are indicated by the background colors. To unravel
different neutrino contributions, nucleosynthesis calculations
considering 
only n.c. or c.c. reactions were performed in addition to
calculations that considered all reactions.

In the following paragraphs, the results for the six nuclei $^7$Li,$^{11}$B,$^{15}$N,$^{19}$F, $^{138}$La
and $^{180}$Ta are discussed.
%7Li-----------------------------------

Without neutrino interactions, the stellar yield of $^7$Li is tiny (0.2
$\%$ of the solar content). Neutrino nucleosynthesis enhances it to 8
$\%$, which still implies, that the solar $^7$Li abundance is produced by a different process,
i.e. by cosmic ray irradiation \cite{Prantzos:2012,Austin.West.Heger:2014}.
The main neutrino mechanism is triggered by n.c. spallation of
protons and neutrons from $^4$He in the stellar He-shell, followed by an 
alpha-capture on the remaining $^3$H and $^3$He fragments.
%11B-----------------------------------
    
    The bulk of $^{11}$B is produced in the thin carbon shell by either
neutrino-induced spallation of protons and neutrons on $^{12}$C or by
$^{12}$C($\nu_e,e^- p)^{11}$C and $^{12}$C(${\bar \nu_e},e^+ n)^{11}$B
reactions, where the $^{11}$C fragments decay with a halflife of about 20
minutes to $^{11}$B. With modern neutrino spectra both n.c. and
c.c. reactions contribute about the same amount. A minor amount of
$^{11}$B is also produced in the O/Ne layer by the $^{16}O(\nu, \nu' \alpha
p)$ reaction, which shows the relative importance of multi-particle emission
reactions. We note that neutrino nucleosynthesis does not produce $^{10}$B in
noticeable amounts. To reproduce the solar $^{11}$B/$^{10}$B ratio of about 4,
Austin et al. \cite{Austin.West.Heger:2014} have argued that neutrino
nucleosynthesis must produce about $(42 \pm 4) \%$ of the solar $^{11}$B
abundance, with the rest stemming from cosmic ray irradiation \cite{Prantzos:2012}.
%15N-----------------------------------

Neutrino nucleosynthesis contributes only a small amount to the solar
$^{15}$N abundance. While n.c. reactions (neutrino-induced
spallation of protons and neutrons from $^{16}$O in the O/C shell) contributed
nearly $10\,\%$ of the solar $^{15}$N yield when the previous neutrino spectra
with higher energies were employed, the modern spectra reduced this amount
significantly.  With these spectra, both neutral- and charged-current reactions
contribute about the same amount. Neutrino nucleosynthesis increases the
solar $^{15}$N abundance by about $3\,\%$. Massive stars and novae are production
sites of $^{15}$N
\cite{Bojazi.Meyer:2014,Jose.Hernanz:2007,Liu.Nittler.ea:2016}.
%19F-----------------------------------

The $\nu$-process enhances the core-collapse supernova
contribution to the solar $^{19}$F abundance by about $30 \%$.  In contrast to
the other nuclides, $^{19}$F shows a distinct sensitivity to the
progenitor mass. For smaller mass stars the hot environment associated with the
passage of the shock wave triggers a thermonuclear sequence,
$^{18}$O$(p,\alpha)^{15}$N$(\alpha,\gamma$)$^{19}$F, in the inner He-shell,
which boosts the pre-supernova amount of $^{19}$F by about $30 \%$, without
inclusion of neutrino reactions.  Neutrino nucleosynthesis by neutral- and
charged-current reactions on $^{20}$Ne in the O/Ne layer adds another $25
\%$.
The thermonuclear component to the $^{19}$F
production is quite sensitive to various nuclear reaction rates, the explosion
energy, and to the compositional shell interfaces. In particular, the mass
contained in the O/Ne layer increases with progenitor mass, and this, in turn,
increases the effectiveness of neutrino nucleosynthesis to $^{19}$F production.
For stars more massive than 17 solar mass, the $\nu$-process can increase the production of
$^{19}$F by factors of $1.5 - 2$.
The main galactic sources of $^{19}$F are Asymptotic Giant Branch stars and
Wolf-Rayet stars
\cite{Renda.Fenner.ea:2004,Kobayashi.Izutani.ea:2011,Cristallo.Leva.ea:2014}.
%138La-----------------------------------

$^{138}$La is a $p$-nucleus, that is bypassed by the $s$-process moving
along the chain of stable barium isotopes. In core-collapse supernovae
the production site is the O/Ne layer where two processes
can occur, both, with and without neutrinos.  At the bottom of the layer,
where temperatures are sufficiently high, $^{138}$La is synthesized by
the $\gamma$-process.  When the peak temperature
drops below $2\,\mathrm{GK}$, photodissociation is less effective and in this cooler region of the O/Ne layer,
neutrino nucleosynthesis dominates the production of $^{138}$La, mainly
by c.c. $(\nu_e,e^-)$ reaction on pre-supernova $^{138}$Ba.  Note
that the $\nu_e$-induced reaction on $^{138}$Ba has also a branch towards
excited states in $^{138}$La above the neutron threshold, that finally lead to
$^{137}$La, whose pre-supernova abundance is tiny. Neutron capture on
$^{137}$La adds then to the production of $^{138}$La.
%180Ta -----------------------------------

$^{180}$Ta is the rarest isotope in Nature.  Its production by
photodissociation and neutrino nucleosynthesis is similar to the production
of $^{138}$La, but with important additional twists. At the high
temperatures in the inner part of the O/Ne layer the $\gamma$-process
produces $^{180}$Ta from pre-supernova $^{181}$Ta. In the cooler outer region of the
O/Ne shell, neutrino nucleosynthesis again dominates the production of $^{180}$Ta by
$(\nu_e,e^-$) reaction on pre-supernova $^{180}$Hf.
Neutral-current reactions contribute in an indirect way as neutron spallation
from $^{16}$O, $^{20}$Ne and $^{24}$Mg generate free neutrons which are
captured on $^{179}$Ta, which is more abundant than
$^{180}$Ta in a low-mass progenitors because of the operation of the $\gamma$-process during the final burning stages. This is the case for
the model shown in Fig. 
\ref{fig:all_profile}, where n.c. and c.c. reactions contribute similarly to the production of  $^{180}$Ta.

$^{180}$Ta is the only isotope in the solar abundance which exists due to a
long-lived isomeric state (with excitation energy of 75 keV), while the ground
state decays with a half-life of about 8 hours.  In the hot environment of the
O/Ne layer, the $^{180}$Ta ground and isomeric states are in thermal
equilibrium. It has been estimated that about $35-39 \%$ of $^{180}$Ta survives
in the isomeric state \cite{Mohr.Kaeppeler.Gallino:2007,Hayakawa.Mohr.ea:2010}.
The production factors of $^{180}$Ta given in Table \ref{tab:prodfac} have been corrected using this estimate.

Two observations are commonly made  regarding the neutrino nucleosynthesis production of
$^{138}$La and $^{180}$Ta. Both isotopes are mainly produced by $(\nu_e,e^-$)
reactions and hence are a tempting way to explore the supernova
$\nu_e$ spectrum (see below). Furthermore, the neutrino production of both
isotopes increases with the size of the O/Ne layer which grows with progenitor
mass. However, this is partly counterbalanced by decreasing abundances from
photodissociation.

\section{Sensitivity to neutrino luminosity and spectra}

Until very recently it has been assumed that the neutrinos, which
trigger the neutrino nucleosynthesis, are generated by cooling of the
proto-neutron star. For all neutrino flavors, their luminosities
and spectra have been approximated by an exponentially decreasing function and
by Fermi-Dirac distributions with zero chemical potential. The 
neutrino temperatures (or average energies) have been individually 
adjusted to supernova simulations. As we have discussed in the last section,
steadily improved descriptions of neutrino opacities within the
simulations led to lower values of the neutrino temperatures that in turn
reduced the abundances of isotopes produced by the $\nu$-process. 
\begin{figure}
 \includegraphics[width=\linewidth]{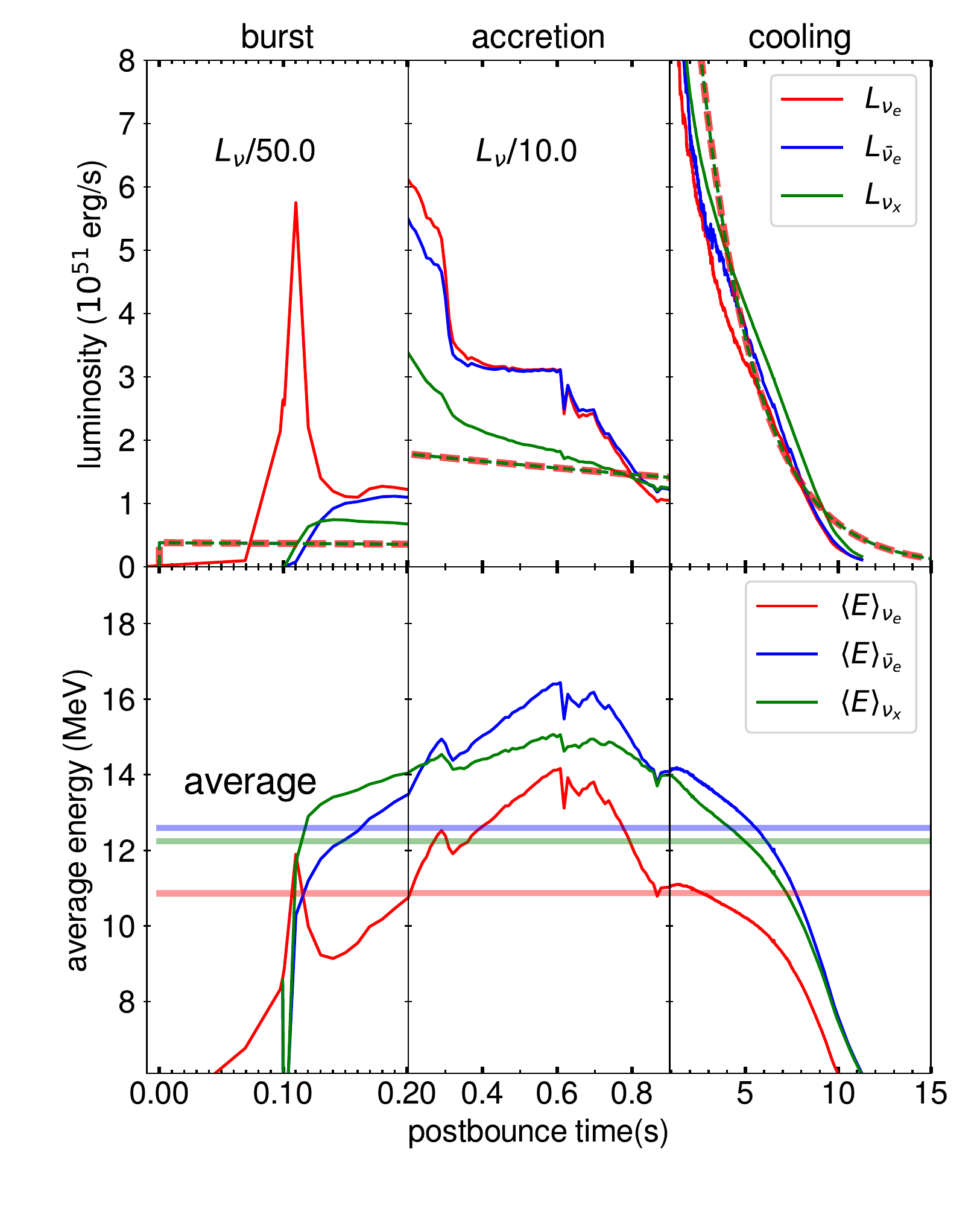}
 \caption{Detailed neutrino signal from a modern supernova simulation \cite{Mirizzi.Tamborra.ea:2016}. Dashed lines in the top panel show the 
 luminosities according to Eq. \ref{eq:lum}, which is appropriate for the cooling phase.
 Horizontal lines in the bottom panel illustrate the 
 time averaged energies that result in the same total neutrino number and energy, when assumed to be constant. The actual average energies substantially 
 exceed these average values, enhancing the effectiveness of the $\nu$-process. }
 \label{fig:signal}
\end{figure}

Fig. \ref{fig:signal} shows the neutrino luminosities and average neutrino energies from a modern
(one-dimensional) supernova simulation \cite{Mirizzi.Tamborra.ea:2016}. The spectra adopted
in neutrino nucleosynthesis studies approximate the cooling phase, which
describes the neutrino emission at a time about $1$~second after bounce. In
this phase, the luminosities drop approximately exponentially, but
also the individual neutrino average energies decrease with time. However,
before the cooling phase dominates the neutrino emission, there is a distinct
burst in $\nu_e$ neutrinos and a so-called accretion phase
with distinct neutrino emission. 
The burst neutrinos are generated from the fast capture of
electrons on free protons that are created after matter 
has been dissociated 
by the shock wave. Before shock revival,
matter falls through the stalled shock, accompanied by
the emission of neutrinos, where in particular $\nu_e$ and $\bar{\nu}_e$
neutrinos have enlarged energies (with respect to the assumed
average value) and luminosities. As stated above, the burst
is in electron neutrinos only. Although it lasts only for about 10 ms,
its luminosity is so large that it carries about $10 \%$
of the total luminosity carried away by electron neutrinos.

Ref. \cite{Sieverding19} has explored which effect the explicit time dependence of the
neutrino luminosities and spectra have on the neutrino nucleosynthesis
yields. The luminosities and average energies were taken from a
recent supernova simulation for a 27~M$_\odot$ progenitor model. For the most
important neutrino-nucleus reactions on $^4$He, $^{12}$C, $^{20}$Ne as well as on $^{138}$Ba and $^{180}$Ta, 
that are discussed in the previous section, the deviations of the spectrum from 
a Fermi-Dirac distribution with vanishing chemical potential, the pinching of the spectra
\cite{Janka.Hillebrandt.ea:1989,Giovanoni.Ellison.ea:1989,Myra.Burrows.ea:1990,Keil.Raffelt.ea:2003},
has also been taken into account,
applying the $\alpha$-fit description of Ref. \cite{Tamborra.Mueller.ea:2012}.

The calculation considering full time dependence of the neutrino
emission has been compared with calculations assuming constant average
neutrino energies, derived as the appropriate mean of the emission
spectrum, conserving total number and energy of each neutrino flavor. The results for $\bar{\nu}_e$ and $\nu_x$ neutrinos 
($T_{\bar{\nu}_e}=4.01$~MeV, $T_{\nu_x}=3.72$~MeV and 
$T_{\bar{\nu}_x} = 3.96$~MeV) are quite
similar to the values adopted in the neutrino nucleosynthesis study
discussed in the previous section. Due to the contributions of the burst
and accretion phases, the mean temperature for electron neutrinos
($T_{\nu_e}=3.46$~MeV) is somewhat larger than the value adopted in \S\ref{sec:nunucl}.

The impact of the time dependence of the neutrino spectra and
luminosities on the neutrino  nucleosynthesis yields is significant (see last column of Table \ref{tab:prodfac}).
The effect is largest for nuclides like $^{138}$La, $^{180}$Ta and 
also $^{11}$B, which are
strongly produced by charged-current reactions and hence are affected
by the modifications of the electron neutrino emissions.
In the calculation that considers the full time dependence,
The enhancement
is larger 
than in calculations with constant average neutrino energies. This has two reasons.
First, the energy dependence of the nuclear cross-sections gives stronger
weight to the early high-temperature neutrinos. Second, in the calculation
with full time dependent spectra, the fluence of
neutrinos through particular stellar mass shells is  larger for the late
neutrinos, which have lower temperatures than the average value. Hence,
a larger number of neutrinos is required to reach the same luminosity value.

The calculation of Ref. \cite{Sieverding19} clearly indicates 
that neutrino nucleosynthesis
studies should consider the full time dependence of neutrino emission,
extended to the full range of stellar masses of core-collapse supernovae.
Furthermore, the duration of the accretion phase is still rather uncertain.
This can also have a significant effect on the $\nu$-process yields,
in particular on nuclei produced by charged-current reactions.       
  
\section{Summary and outlook}

It is by now well established that the solar abundances of 
$^7$Li, $^{11}$B, $^{19}$F,
$^{138}$La and $^{180}$Ta are, to an important or even dominating part, being 
produced by the $\nu$-process. Nevertheless, increasingly realistic studies
with improved stellar models, better nuclear cross-sections  and also
improved neutrino emission data from advanced supernova simulations
show that the individual abundances of these nuclides depend sensitively 
on astrophysical and nuclear modeling. The largest impact on
the $\nu$-process yields has been due to changes in the neutrino spectra
that are predicted with significantly lower average energies by modern
supernova simulations than anticipated in earlier studies. This spectra
change is reflected in the noticeable reduction of the various abundances.
However, a recent study put in question the treatment of neutrino emission that
assumes time-independent neutrino spectra. In fact, larger
yields are found if the time dependence of neutrino emission is explicitly
accounted for. This is particularly true for the charged-current contributions
arising mainly from electron neutrinos from the early burst and accretion phases
of the neutrino emission. Realizing this sensitivity, two steps are recommended to encourage advancement in this field:
the study, performed for a 27~$M_\odot$ progenitor,
must be extended to the full range of core-collapse supernova progenitor
masses and neutrino nucleosynthesis should be extended from one-dimensional
supernova models, to which they were restricted recently, to 
multi-dimensional models.

Early on it has been realized that neutrino nucleosynthesis is
sensitive to those neutrino flavors ($\nu_e, \nu_x$) which have not
been observed from supernova SN1987A (whose signal was likely due
to electron anti-neutrinos). Here $^{138}$La and $^{180}$Ba, which are 
dominantly made by charged-current reactions, serve as a constraint for
the $\nu_e$ spectra and luminosities, while the other $\nu$-process
nuclides are also sensitive to neutral-current reactions and thus
to the other neutrino flavor spectra. It has been even pointed out
that neutrino nucleosynthesis may be critical in determining the neutrino
mass hierarchy or the famous $\theta_{13}$ angle of the neutrino 
matrix \cite{Yoshida.Terasawa.ea:2004,Yoshida.Kajino.ea:2005,Yoshida.Kajino.ea:2006,Yoshida.Suzuki.ea:2008,Mathews.Kajino.ea:2012,Kajino.Mathews.ea:2014}.
The clue here is always that neutrino nucleosynthesis is expected
to be sensitive to neutrino oscillations that shuffle the lower-energy
$\nu_e$ spectrum with the higher-energy spectra of muon or tau neutrinos. 
The more complex effects of collective neutrino oscillations can also have an
impact \cite{Wu.Qian.ea:2015}.
While this observation is in principle correct, a deduction of the
important neutrino properties from $\nu$-process yields certainly
becomes more difficult as the energy hierarchy of the supernova neutrinos
is less pronounced in modern supernova simulations and further difficulties arise when
the time dependence of the neutrino emission is considered. 

\begin{acknowledgments}
  This work has been partly supported by the Deutsche
  Forschungsgemeinschaft through contract SFB~1245
\end{acknowledgments}

%

%\bibliography{bib}

\end{document}